\documentclass[floats,aps,showpacs,twocolumn,superscriptaddress]{revtex4}

\usepackage{amsmath,bm}
\usepackage[dvips]{graphicx} 
\DeclareGraphicsExtensions{.eps}

\begin{document}

\title{Noise properties of  a resonance-type  spin-torque microwave detector}

\author{Oleksandr Prokopenko}
\email{ovp@univ.kiev.ua}
\author{Gennadiy Melkov}
\affiliation{Faculty of Radiophysics, Taras
Shevchenko National University of Kyiv, Kyiv 01601, Ukraine}

\author{Elena Bankowski}
\author{Thomas Meitzler}
\affiliation{U.S. Army TARDEC, Warren, MI 48397, USA}

\author{Vasil Tiberkevich}
\author{Andrei Slavin}
\affiliation{Department of Physics, Oakland University, Rochester,
MI 48309, USA}

\begin{abstract}
We analyze performance of a resonance-type  spin-torque microwave
detector (STMD) in the presence of noise and reveal two distinct
regimes of STMD operation. In the first (high-frequency) regime the
minimum detectable microwave power $P_{\rm min}$ is limited by the
low-frequency Johnson-Nyquist noise and the signal-to-noise ratio
(SNR) of STMD is proportional to the input microwave power $P_{\rm
RF}$. In the second (low-frequency) regime $P_{\rm min}$ is limited
by the magnetic noise, and the SNR is proportional to $\sqrt{P_{\rm
RF}}$.  The developed formalism can be used for the optimization of
the practical noise-handling parameters of a STMD.
\end{abstract}


\date\today

\maketitle

It has been shown in \cite{tulapurkar05a, suzuki10b} that a magnetic
tunnel junction (MTJ) subjected to the external microwave current
$I_{\rm RF}(t) = I_{\rm RF} \cos(2 \pi f t)$ can perform as
a resonance-type quadratic detector of microwave radiation
generating the DC voltage $U_{\rm DC}$ proportional to the acting
microwave power $U_{\rm DC} =\varepsilon P_{\rm RF}$ ($P_{\rm RF}
\sim I_{\rm RF}^2$). The detector operation is based on the
spin-torque effect \cite{slonczewski96a,berger96a} and the detector
sensitivity $\varepsilon$ has a maximum value
$\varepsilon=\varepsilon_{\rm res}$ when the frequency of the
external microwave signal is close to the eigen-frequency $f_0$ of
the MTJ nanopillar, $f = f_0$.

The resonance sensitivity $\varepsilon_{\rm res}$ of the spin-torque
microwave detector (STMD) was calculated in \cite{wang09a}:
\begin{equation}
\label{Sensitivity}
    \varepsilon_{\rm res}
    =
    \frac{U_{\rm DC}}{P_{\rm RF}}
    =
    \left(\frac{\gamma \hbar}{4 e}\right)
    \frac{P^3}{M_s V \Gamma}
    Q(\theta_0) \ ,
\end{equation}
where $\gamma \approx 2\pi \cdot 28 \,{\rm GHz}/{\rm T}$ is the
modulus of the gyromagnetic ratio, $\hbar$ is the reduced Planck
constant, $e$ is the modulus of the electron charge, $P$ is the spin-polarization
efficiency of the MTJ, $M_s$ is the
saturation magnetization of the free layer (FL) of MTJ, $V = \pi r^2 d$ is the volume of the FL ($r$
is its radius and $d$ is its thickness),  $\Gamma$ is the magnetization
damping rate in the MTJ FL proportional to the Gilbert damping
constant $\alpha$, and $Q(\theta_0)$ is the geometrical factor that depends on the angle $\theta_0$ between the directions
of the equilibrium magnetization in FL and pinned layer (PL) of the MTJ. For an in-plane magnetized MTJ, $Q(\theta_0) =
\sin^2\theta_0 / (1 + P^2\cos\theta_0)^2$.

Estimations based on Eq.~(\ref{Sensitivity}) \cite{wang09a} and recent experimental results \cite{suzuki10b, krivorotov10a} have
demonstrated that the STMD sensitivity can exceed that of passive
semiconductor Schottky-diode microwave detectors ($\varepsilon \sim 1000 \,{\rm
V/W}$), which makes STMD very interesting for practical applications
in microwave measurement technology.

The operation and the minimum detectable power of all types of
microwave detectors are limited by noise (in particular, by the
low-frequency Johnson-Nyquist (JN) noise in the case of unbiased
Schottky diodes \cite{diode_noise}), and, therefore, it is important
to understand the noise-handling properties of the novel STMD based
on MTJ.

In this Letter we present theoretical analysis of the noise
properties of a passive  STMD (no DC bias current) using the STMD
model developed in \cite{wang09a} with additional terms describing influence of
thermal fluctuations. In our analysis we took into account three
sources of noise:
\par (a) Low-frequency Johnson-Nyquist (JN) noise voltage $U_{\rm N}(t)$
associated with the MTJ resistance $R_0$. This type of
noise is additive and is independent of the magnetization dynamics.
\par(b) High-frequency JN noise current $I_{\rm N}(t)$
which transforms into a non-additive low-frequency noise after
mixing with the microwave oscillations caused
by the input microwave signal.
\par(c) Magnetic noise (MN), which is caused by the thermal fluctuations
of the magnetization direction in the MTJ FL. This noise, modeled by a random magnetic field $\bm B_{\rm N}(t)$, leads to the
fluctuations of the electrical resistance of the STMD and transforms to low frequencies
after mixing with the driving current $I_{\rm RF}(t)$.

The dynamics of magnetization ${\bm M}$ in the MTJ FL under the
action of a microwave current containing both deterministic and
noise components $I(t)=I_{\rm RF}(t) + I_{\rm N}(t)$ and noise magnetic
field ${\bm B}_{\rm N}(t)$ is described by the
Landau-Lifshits-Gilbert-Slonczewski equation:
\begin{eqnarray}
\label{LLGS}
    \frac{d\bm M}{dt} &=& \gamma[\bm B_{\rm eff}(\bm M) \times \bm M]
            + \frac{\alpha}{M_s}\left[\bm M \times \frac{d\bm M}{dt}\right]+
\nonumber\\&&
            \frac{\sigma I(t)}{M_s}[\bm M \times [\bm M \times \bm p]]
            + \gamma[{\bm B}_{\rm N}(t) \times \bm M]
\ ,
\end{eqnarray}
where $\bm B_{\rm eff}(\bm M)$ is the effective magnetic field,
which includes the external bias magnetic field ${\bm
B}_0$ and the demagnetization field,
$\bm p$ is the unit vector in the direction of the magnetization
of the MTJ PL, $\sigma = (\gamma\hbar/2e)P/[(1 + P^2\cos\theta)
M_s V]$ is the current-torque proportionality coefficient, and $\theta$ is the angle between vectors $\bm M$ and $\bm p$.

We performed analysis for a ``planar'' STMD, in which both FL and PL are magnetized in-plane. In this case, the MTJ eigen-frequency is $f_{\rm 0}
= (\gamma/2\pi)\sqrt{B_0(B_0 + \mu_0 M_s)}$ and the damping rate has the form $\Gamma = \alpha\gamma(B_0 + \mu_0 M_s/2)$,
where $\mu_0$ is the vacuum susceptibility. We assumed that $f = f_0 \gg \Gamma/(2\pi)$. Using Eq.~(\ref{LLGS}), we found the linear FL magnetization response to current $I(t)$ and field $B_{\rm N}(t)$ in the frequency domain. The output STMD signal was calculated as the low-frequency part of the voltage $[R(\theta)I(t) + U_{\rm N}(t)]$, where $R(\theta) = R_\bot/(1+P^2\cos\theta)$ is the MTJ magnetoresistance, $R_{\bot} = {\rm RA}/(\pi r^2)$ is the
MTJ resistance in the perpendicular magnetic state ($\theta = \pi/2$), and ${\rm RA}$ is the resistance-area product of the MTJ.
We assumed that all noise sources are independent Gaussian processes with uniform spectral densities $S(U_{\rm N}) = S(I_{\rm N})R_0^2 = 2 k_B T R_0$ and $S(B_{\rm N}) = 2 \alpha k_B T/(\gamma M_s V)$, where $k_B$ is the Boltzmann constant, $T$ is the noise
temperature, and $R_0 = R(\theta_0)$ is the equilibrium resistance of MTJ. The detailed derivation of the noise spectrum of STMD will be presented elsewhere \cite{unpublished}.

We found that the noise of the output voltage $U_{\rm DC} = \varepsilon_{\rm res} P_{\rm RF}$ has characteristic spectral width of $\Gamma$ and, for typical frequency bandwidth of measurement $\Delta f \ll \Gamma/(2\pi)$, can be considered as frequency-independent. The root mean square fluctuations $\Delta U_{\rm DC}$ can be written in the simple form
\begin{equation}
\label{dUdc}
    \Delta U_{\rm DC} = U_{\rm JN} \sqrt{1 + \frac{U_{\rm DC}}{U_{\rm IM}} + \frac{U_{\rm DC}}{U_{\rm MN}}} \ ,
\end{equation}
where the three terms in the right-hand side part of the equation
describe, respectively, the influence of the three above mentioned
noise sources and
\begin{subequations}
\label{dUterms}
\begin{eqnarray}
\label{dUterms-a}
    U_{\rm JN} &=& \sqrt{4 k_B T R_0 \Delta f} \ , \\
\label{dUterms-b}
    U_{\rm IM} &=& \frac{R_0}{4 \varepsilon_{\rm res}} \ , \
    U_{\rm MN} = \frac{\gamma\hbar}{2e}\frac{B_0}{P} \ .
\end{eqnarray}
\end{subequations}

For typical parameters (see e.g. \cite{wang09a,suzuki10a}) of an
MTJ nanopillar ($r = 50 \, {\rm nm}$, $d = 1 \, {\rm nm}$, $P = 0.7$,
$\theta_0 = 45 \ {\rm deg}$, $R_0 = 500 \, \Omega$
(giving ${\rm RA} = 5.29 \, \Omega \,\mu{\rm m}^2$), $\alpha = 0.01$,
$\mu_0 M_s = 0.8 \, {\rm T}$, $B_0 = 38 \, {\rm mT}$ (giving $f_0
= 5 \, {\rm GHz}$)) Eq.~(\ref{Sensitivity}) gives the
resonance STMD sensitivity in the passive regime $\varepsilon_{\rm
res} \approx 700 \, {\rm V/W}$, which is comparable to the
sensitivity of Schottky diodes \cite{suzuki10a}. At room temperature
$T = 300 \ {\rm K}$ and for the measurement bandwidth of $\Delta f =
1 \ {\rm MHz}$ Eqs.~(\ref{dUterms}) give the following estimations
for the noise-induced voltages: $U_{\rm JN} = 2.88 \, \mu{\rm V}$,
$U_{\rm IM} = 175.9 \, {\rm mV}$, $U_{\rm MN} = 3.14 \, \mu{\rm V}$.

It is clear that the characteristic voltage $U_{\rm IM}$ of the
non-additive high frequency JN noise is much larger than the
voltages created by the low frequency JN noise $U_{\rm JN}$,
magnetic noise $U_{\rm MN}$, and the typical DC voltage output
$U_{\rm DC}$ of the STMD. This means that the influence of the
high-frequency JN noise in Eq.~(\ref{dUdc}) can be completely
ignored. Note, also, that, in contrast with the other characteristic
noise voltages, the voltage $U_{\rm MN}$ caused by the magnetic
noise is proportional to the bias magnetic field $B_0$, and,
therefore, increases with the increase of the frequency of the input
microwave signal.

Now, introducing the microwave powers $P_{\rm RF} = U_{\rm
DC}/\varepsilon_{\rm res}$,  $P_{\rm JN} = U_{\rm
JN}/\varepsilon_{\rm res}$, $P_{\rm MN} = U_{\rm
MN}/\varepsilon_{\rm res}$ and using Eq.~(\ref{dUdc}), we can write
a simple expression for the signal-to-noise ratio (SNR) of the STMD
in terms of these characteristic powers:
\begin{equation}
\label{SNR}
    {\rm SNR}
    =
    \frac
    {
        U_{\rm DC}
    }
    {
        \Delta U_{\rm DC}
    }
    =
    \frac{ P_{\rm RF} }{ P_{\rm JN} }
    \sqrt
    {
        \frac{ P_{\rm MN} }{ P_{\rm MN} + P_{\rm RF} }
    } \ .
\end{equation}

The simple analysis of Eq.~(\ref{SNR}) demonstrates that there are
two distinct regimes of operation of the resonance STMD in the
presence of thermal noise. We shall classify them by the type of
noise that limits the minimum detectable power of STMD $P_{\rm min}$ (power
corresponding to ${\rm SNR}=1$).

The first regime corresponds to the case of relatively high
frequencies of the input microwave signal, when $P_{\rm MN} \gg P_{\rm RF}$ (for $P_{\rm RF} \sim P_{\rm min}$).
In this regime, similar to the conventional semiconductor
diodes, the minimum detectable power is limited by the
low-frequency JN noise, $P_{\rm min} = P_{\rm JN}$, and
 the SNR of STMD is linearly proportional
to the input microwave power $P_{\rm RF}$ (${\rm SNR} \simeq P_{\rm
RF}/P_{\rm JN}$).

The second regime takes place in the opposite limiting case of
relatively low input frequencies, when  $P_{\rm MN} \ll P_{\rm RF}$.
In this case the SNR of STMD increases with $P_{\rm RF}$ much
slower than in conventional diodes, and is proportional to the
square root of the input microwave power: ${\rm SNR} \simeq
\sqrt{P_{\rm RF}/P_{\rm min}}$. The minimum detectable power $P_{\rm
min} = P^2_{\rm JN}/P_{\rm MN}$ in this regime is limited by the
magnetic noise in the FL of the MTJ.

\begin{figure}
\includegraphics{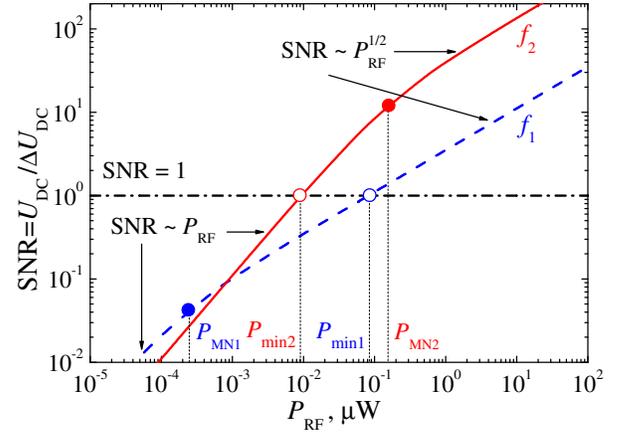}
\caption { Dependence of the SNR of STMD on the input microwave power $P_{\rm RF}$
calculated from Eq.~(\ref{SNR}) for two different frequencies of the
input microwave signal: $f_1 = 1 \, {\rm GHz}$ (dashed blue
line) and $f_2 = 25 \, {\rm GHz}$ (solid red line).
All other parameters are the same as indicated below Eqs.~(\ref{dUterms}).
$P_{\rm
min}$ is the minimum detectable power of STMD (at ${\rm SNR}=1$) and $P_{\rm MN}$ is
 the frequency-dependent characteristic power of magnetic noise. }
\label{FigSNR@PRF}
\end{figure}

We note, that if at a fixed frequency of the input signal the SNR of
an STMD is measured in a wide range of input powers, covering both
the above described limiting cases, it is possible to find the
characteristic power of the magnetic noise $P_{\rm MN}$ with a good
accuracy, and, using expression (\ref{dUterms-b}) for $U_{\rm MN}$,
to determine with the same accuracy the MTJ spin-polarization
efficiency $P$.

The existence of two distinct regimes of STMD operation is
illustrated in Fig.~\ref{FigSNR@PRF}, where two curves calculated
from Eq.~(\ref{SNR}) show the STMD SNR as functions of the input
power for signal frequencies $f_1$ = 1~GHz (dashed blue line)
and $f_2$ = 25~GHz (red solid curve). It can be seen, that
both curves (presented in logarithmic coordinates) demonstrate the
clear change of slope from 1  to 1/2  in the region, where the input
power $P_{\rm RF}$ is close to the characteristic power of the
magnetic noise $P_{\rm MN}$ (which increases with the increase of
the input signal frequency). The minimum detectable power $P_{\rm
min}$(corresponding to ${\rm SNR}=1$) in the high-frequency case is smaller
than $P_{\rm MN}$ and lies in the region of the linear dependence of
SNR on $P_{\rm RF}$ (solid red line in Fig.~\ref{FigSNR@PRF}). The
situation is opposite in the low frequency case (blue dashed curve
in Fig.~\ref{FigSNR@PRF}), when $P_{\rm
min} > P_{\rm MN}$ and lies in the region, where the slope of the
SNR curve is equal to 1/2.

\begin{figure}
\includegraphics{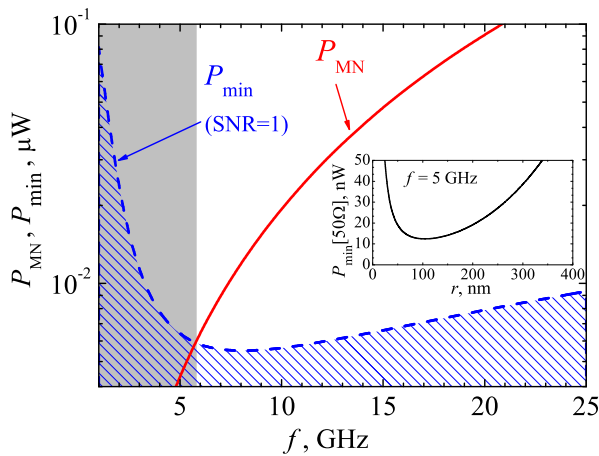}
\caption { Characteristic power of magnetic noise $P_{\rm MN}$
(solid red line) and minimum detectable power $P_{\rm min}$ of STMD
(dashed blue line) as functions of the input microwave frequency
$f$. The blue dashed area corresponds to undetectable signals
$P_{\rm RF} < P_{\rm min}$ and gray shaded area shows the low-frequency
STMD regime, where the magnetic noise is dominant in the whole practical
region $P_{\rm RF} > P_{\rm min}$.
Inset: minimum detectible microwave power delivered to
a 50-$\Omega$ transmission line $P_{\rm min}[50 \, \Omega]$ for
$f$ = 5~GHz as a function of the radius $r$ of the MTJ
nanopillar.
STMD parameters are indicated below Eqs.~(\ref{dUterms}).
} \label{FigPminPMNF}
\end{figure}

The evolution of the characteristic powers $P_{\rm MN}$ and $P_{\rm
min}$ with the increase of frequency of the input microwave signal
is shown in Fig.~\ref{FigPminPMNF}. The curve $P_{\rm
MN}(f)$ separates the plane into the region, where magnetic noise is dominant (above the
curve), and the region, where the STMD operation is limited by the JN
noise (below the curve). It is, clear, that the smallest detectable
power is achieved near the border of these two regimes.

When an STMD based on an MTJ nanopillar is used as a sensor of microwave
radiation, it is typically connected to a standard transmission line
with the impedance of $Z_{\rm TL} = 50 \,\Omega$.  The minimum
detectable microwave power {\em delivered to a 50-$\Omega$
transmission line} can be written as \cite{pozar05a} $P_{\rm min}[50 \, \Omega] =
(1/4)(R_0 + Z_{\rm TL})^2 P_{\rm min}/Z_{\rm TL} R_0$.
Using this expression and taking into account the size dependence of the
STMD resistance ($R_0 \propto 1/r^2$), it is possible to show that $P_{\rm min}[50 \,
\Omega]$ has a clear minimum as a function of the nanopillar radius
$r$.  For instance, the optimum value of the nanopillar radius is
$r_{\rm opt} \approx 100 \,{\rm nm}$ for the input frequency
$f = 5 \,{\rm GHz}$ (see the inset in Fig.~\ref{FigPminPMNF}).

In conclusion, we have demonstrated that STMD in the presence of
noise can operate in two distinct regimes, one of which is limited
by magnetic noise and is  different from the regime of operation of
traditional semiconductor detectors. We have, also, suggested that
the measurements of STMD SNR in a wide range of input powers can be
used to determine the spin-polarization efficiency $P$ of MTJ
nanopillars, and have shown that the developed formalism can be used
for the optimization of noise-handling parameters of a STMD.

This work was supported in part by the Contract from the U.S.~Army
TARDEC, RDECOM, by the grants ECCS-1001815 and DMR-1015175 from the
National Science Foundation of the USA, by the Grant No.~M/90-2010
from the Ministry of Education and Science of Ukraine, and by the
Grant No.~UU34/008 from the State Fund for Fundamental Research of
Ukraine.

\end{document}